\documentclass[12pt]{article}

\usepackage{natbib}
\usepackage{epsfig}

\begin{document}

\renewcommand{\refname}{\normalsize \bf \em References}

\title{\bf  COLE-COLE ANALYSIS OF THE SUPERSPIN GLASS SYSTEM
Co$_{80}$Fe$_{20}$/Al$_{2}$O$_{3}$}
\author{
O.\ PETRACIC$^{a,}$\footnote{
      Corresponding author: Tel.\ +49-203-379-2939;
      fax: +49-203-379-1965,
      email: petracic@kleemann.uni-duisburg.de}\ , 
S.\ SAHOO$^a$, CH.\ BINEK$^a$, W.\ KLEEMANN$^a$,
 \\*[0.0cm]
J.\ B.\ SOUSA$^b$, S.\ CARDOSO$^c$ and P.\ P.\ FREITAS$^c$
\\*[0.0cm]
$^a${\small \it Laboratorium f\"ur Angewandte Physik,} \\
    {\small \it Gerhard-Mercator-Universit\"at Duisburg, 47048 Duisburg, Germany}
\\*[0.0cm]
$^b${\small \it IFIMUP, Departamento de Fisica,} \\ 
     {\small \it Universidade de Porto,  
     4169-007 Porto, Portugal}
\\*[0.2cm]
$^c${\small \it INESC,} 
     {\small \it Rua Alves Redol 9-1,  
     \small 1000 Lisbon, Portugal}
}
\date{\small (Received \hspace{3cm} )}
\maketitle

\begin{abstract}

Ac susceptibility measurements were performed on discontinuous magnetic
multilayers
$\lbrack$Co$_{80}$Fe$_{20}(t)$/Al$_{2}$O$_{3}$(3nm)$\rbrack$$_{10}$,
$t = 0.9$ and 1.0~nm,
by Superconducting Quantum Interference Device (SQUID) magnetometry.
The CoFe forms nearly spherical ferromagnetic
single-domain nanoparticles in the
diamagnetic Al$_{2}$O$_{3}$ matrix. Due to dipolar interactions and random
distribution of anisotropy axes the system exhibits a spin-glass phase.
We measured the {\it ac} susceptibility as a function of temperature 
$20 \leq T \leq 100$~K at different {\it dc} fields and as a function of frequency
$0.01 \leq f \leq 1000$~Hz. The spectral data were successfully
analysed by use of the
phenomenological Cole-Cole model, giving a power-law temperature
dependence of the characteristic relaxation time $\tau_c$ and a
high value for the polydispersivity exponent, $\alpha \approx 0.8$,
typical of spin glass systems.

\end{abstract}

\noindent
{\it Keywords}: Multilayers, Ac susceptibility, Polydispersivity, 
Dipolar interactions, Spin glass behavior

\section*{INTRODUCTION}

The dynamic and static magnetic properties of spin glasses (SG) are still
a subject of intense experimental and theoretical research. In the field 
of experiments a vast variety of different spin glass systems 
\citep*{Young_book} have yet been found and investigated. 
A rather new class
are the so-called superspin glass (SSG) systems \citep*{Kleemann_01a}. 
Here the sample is composed
of an ensemble of dipolarly interacting nanoparticles, each 
having a superspin moment in the order of 1000~$\mu_B$. 
The spin glass properties are due to frustration, a natural property
of dipolar interaction, and to randomness of the anisotropy axis
directions, frequently also of the spin sizes.
Two different types of  
realisations of SSG systems exist, frozen ferrofluids 
\citep*{Dormann_97, Dormann_99b, Djurberg_97, Mamiya_99} 
and 
discontinuous magnetic multilayers 
\citep*{Sankar_00, Sousa_01, Kleemann_01a, Petracic_02a, 
Sahoo_02a}.

It is widely accepted that 3-dimensional (3D) nanoparticle systems 
with high
enough density of the particles and sufficiently narrow particle size 
distribution do have spin 
glass properties, 
i.e. there exists a phase transition temperature, $T_g$, where
the characteristic relaxation time and the static non-linear
susceptibility diverge 
\citep*{Djurberg_97, Kleemann_01a, Petracic_02a, Sahoo_02a}.
In order to observe SSG properties the collective glass temperature,  
$T_g$, has to be larger than the so-called blocking temperature, $T_b$,
at which the relaxation time of the individual moments
\citep*{Neel_49, Brown_63} 

\begin{equation} \label{eq-neel}
\tau = \tau_0 \exp(KV/k_BT),
\end{equation}
 
\noindent reaches the order of the timescale of the experiment. Here $K$ 
is the
effective anisotropy constant of one nanoparticle, $V$ the volume and
$\tau_0 \sim 10^{-10}$s the relaxation time at $T \rightarrow \infty$. Below
$T_b$ the particle moments are "blocked". 

The condition $T_g < T_b$ is met in our
discontinuous metal-insulator multilayers (DMIMs) 
$\lbrack$Co$_{80}$Fe$_{20}(t)$/Al$_{2}$O$_{3}$(3nm)$\rbrack$$_{n}$,
where $t \leq 1.0$~nm is the nominal thickness of the ferromagnetic CoFe 
layers and
$n$ the number of bilayers. The CoFe does not form a continuous layer
but forms nearly spherical particles embedded in the diamagnetic
Al$_{2}$O$_{3}$ matrix \citep*{Kakazei_01}. 
One finds self-organized arrangements of
particles in each layer \citep*{Stappert_02}, i.e. the inter-particle
distances are nearly constant. 
While SSG behavior is found for relatively small values of 
$t \leq 1.0$~nm
and $n = 10$ 
\citep*{Kleemann_01a, Petracic_02a, Sahoo_02a},
for higher values of the
nominal thickness, $1.0 < t \leq 1.4$~nm, superferromagnetism (SFM)
is observed 
\citep*{Kleemann_01a, Sahoo_02b, Chen_02a}. 

In this article we will focus on the SSG systems 
$\lbrack$Co$_{80}$Fe$_{20}(t)$/ Al$_{2}$O$_{3}$(3nm)$\rbrack$$_{10}$,
with $t = 0.9$ and $1.0$~nm. The existence of a spin glass phase
was evidenced by means of dynamic criticality, static criticality
of the non-linear susceptibililty and dynamical scaling
\citep*{Kleemann_01a, Petracic_02a, Sahoo_02a}. 
All three methods yield convincing values
for the glass transition temperature and the dynamical critical
exponents, respectively, 
$T_g \approx 44$~K, 
$z \nu \approx 9.5$, $\gamma \approx 1.47$ and 
$\beta \approx 1.0$ for $t = 0.9$~nm and
$T_g \approx 49$~K, 
$z \nu \approx 10.0$, $\gamma \approx 1.36$ and 
$\beta \approx 0.6$ for $t = 1.0$~nm. $T_g$ and $z \nu$
are the error weighted average values obtained
from different methods.

While $z \nu$ characterizes the divergence of the relaxation time of the
largest ordered cluster as $T \rightarrow T_g$, there is a wide
distribution of shorter relaxation times due to non-percolating
clusters. They are characteristic of the glassy nature of the system
and deserve a focused investigation, which will be described in the 
present paper. To this end we analyse the results of measurements
of the complex {\it ac} susceptibility carried out at different
{\it ac} amplitudes and bias fields and frequencies $f$. In particular
the so-called Cole-Cole presentation, $\chi''$ vs $\chi'$, will
be discussed in terms of appropriate empirical models of
relaxational polydispersivity.

\section*{EXPERIMENTAL}

The DMIM samples 
Glass/Al$_{2}$O$_{3}$(3nm)/$\lbrack$Co$_{80}$Fe$_{20}(t)$/
Al$_{2}$O$_{3}$(3nm)$\rbrack$$_{10}$ ($t = 0.9$ and $1.0$~nm) are 
prepared by sequential Xe ion beam sputtering from two seperate 
targets \citep*{Kakazei_01}. The CoFe forms nearly spherical 
granules of 
approximately 3~nm diameter and 2~nm inter-particle spacing
as found from transmission electron
microscopy (TEM) studies \citep*{Stappert_02}. 

The measurements were performed by use of a commercial
Superconducting Quantum Interference Device (SQUID)
magnetometer (MPMS-5S, Quantum Design). The 
{\it ac} susceptibility, $\chi = \chi' -i \chi''$,  is extracted from the 
linear response of the sample on an oscilllating {\it ac} field,
$\mu_0 H_{ac} = 0.05$ or $0.4$~mT at different {\it ac} frequencies,
$0.01 \leq f \leq 1000$~Hz. The constant {\it dc} field was either
$\mu_0 H = (0 \pm 0.03)$~mT or $(0.6 \pm 0.1)$~mT.

\section*{RESULTS AND DISCUSSION}

\begin{figure}[th] \begin{center}
\epsfig{figure=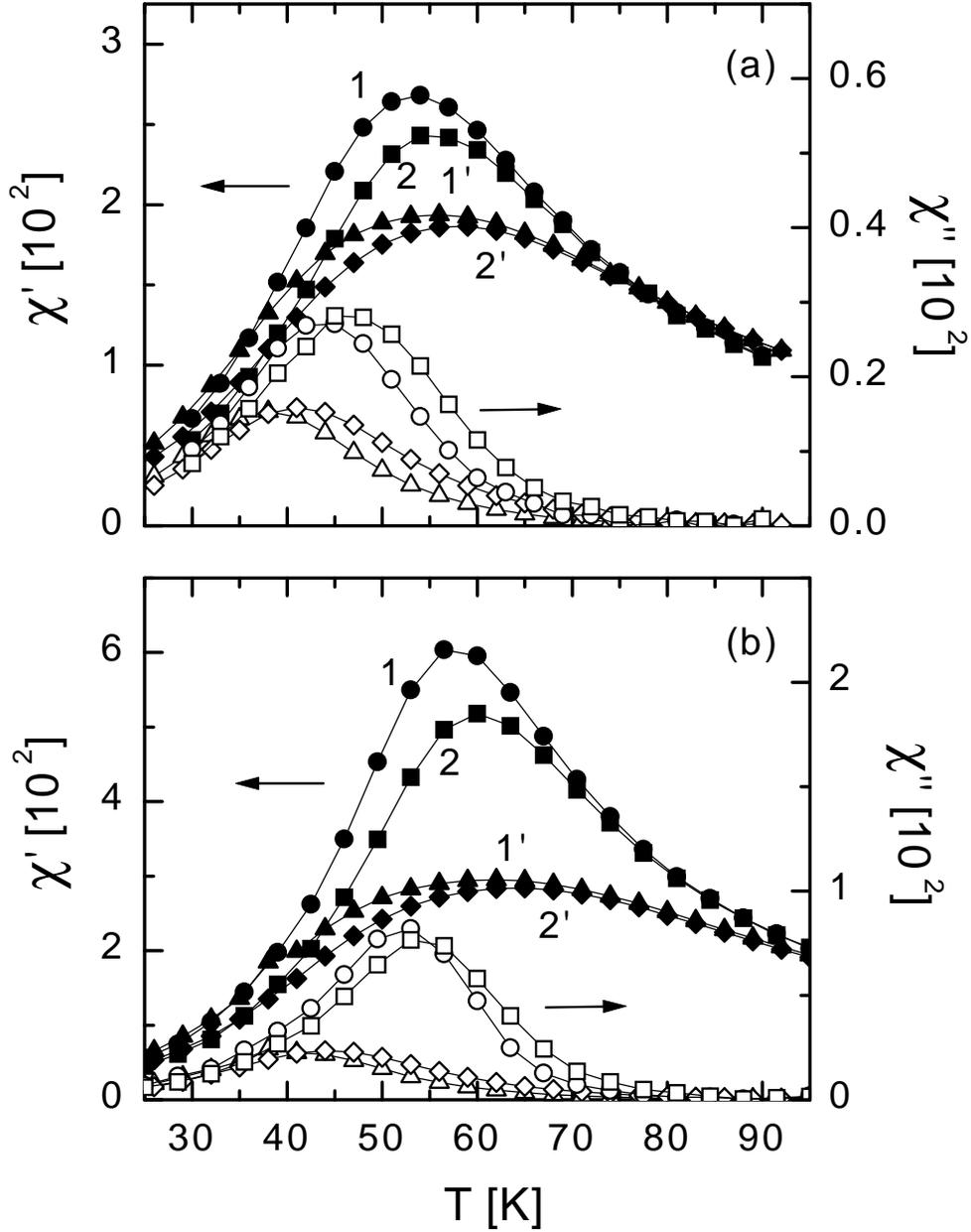, width=13.8cm}
\caption{$\chi'$ and $\chi''$ vs $T$ for $t = 0.9$~nm (a) and 1.0~nm (b) 
measured at constant frequency 
$f = 0.1$ (curves 1 and 1') and 1~Hz (curves 2 and 2') with
{\it ac} field amplitude $\mu_0 H_{ac} = 0.05$~mT and {\it dc} field
$\mu_0 H = 0$~mT (curves 1 and 2) and 
{\it ac} field amplitude $\mu_0 H_{ac} = 0.4$~mT and {\it dc} field
$\mu_0 H = 0.6$~mT (curves 1' and 2'), respectively.
}
\label{fig1}
\end{center} \end{figure}

\begin{figure}[th] \begin{center}
\epsfig{figure=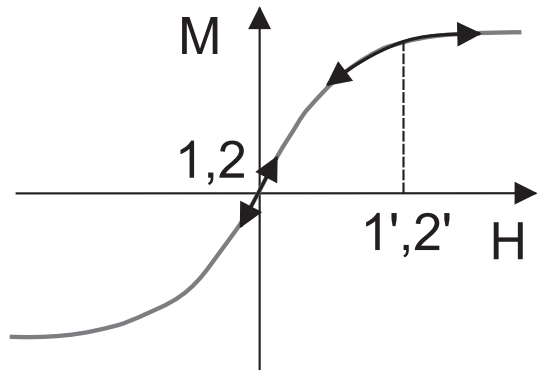, width=7cm}
\caption{Schematic drawing of the measurement conditions relevant
for the data as numbered in Figure 1 (see text). 
The solid curve shows $M(H)$ without hysteresis.
}
\label{fig1b}
\end{center} \end{figure}

Figure~\ref{fig1} shows the real $\chi'$ and the imaginary parts 
$\chi''$ of the {\it ac} susceptibility vs temperature $T$ for the samples 
$t = 0.9$~nm 
(a) and $1.0$~nm (b) under four different conditions. Curves 1 and 1'
are measured at the {\it ac} frequency $f = 0.1$ and curves 2 and 2'
at 1~Hz, whereas for curves 1 and 2 an {\it ac} field amplitude of 
$\mu_0 H_{ac} = 0.05$~mT and a {\it dc} field of $\mu_0 H = 0$~mT were 
applied.
For curves 1' and 2' an {\it ac} field amplitude of $\mu_0 H_{ac} = 
0.4$~mT and a {\it dc} field of $\mu_0 H = 0.6$~mT were used (see 
Figure~\ref{fig1b} for an illustration). For both samples a similar 
behavior is encountered. Both the increase of the probing
{\it ac} field amplitude and the
application of a bias field result
in a supression of the amplitude of the real part $\chi'(T)$ and
a shift $\Delta T_m$ of the peak to higher temperatures.
Quantitatively the shift is $\Delta T_m(1'-1)=2.2$ $(3.8)$~K and 
$\Delta T_m(2'-2)=2.3$ $(3.6)$~K for $t = 0.9$ $(1.0)$~nm, respectively.
The imaginary part $\chi''(T)$ is also suppressed, but the inflection point
at $T_f$ is shifted to lower temperatures,
$\Delta T_f(1'-1)=-6.0$ $(-10.1)$~K and
$\Delta T_f(2'-2)=-6.2$ $(-8.2)$~K for $t = 0.9$ $(1.0)$~nm, respectively.
This behavior is well known from other SG systems and
model calculations 
\citep*{Canella_72, Barbara_81} 
and can be
explained in terms of a competition between the non-critical linear
susceptibility and the critical non-linear susceptibility. In other
words, 
the suppression of both 
the real and the imaginary parts 
reflects the obvious fact that the 
$M(H)$ curve becomes increasingly non-linear
when increasing the {\it ac} amplitude and/or the bias field 
(see Fig.~\ref{fig1b}).

\begin{figure}[th] \begin{center}
\epsfig{figure=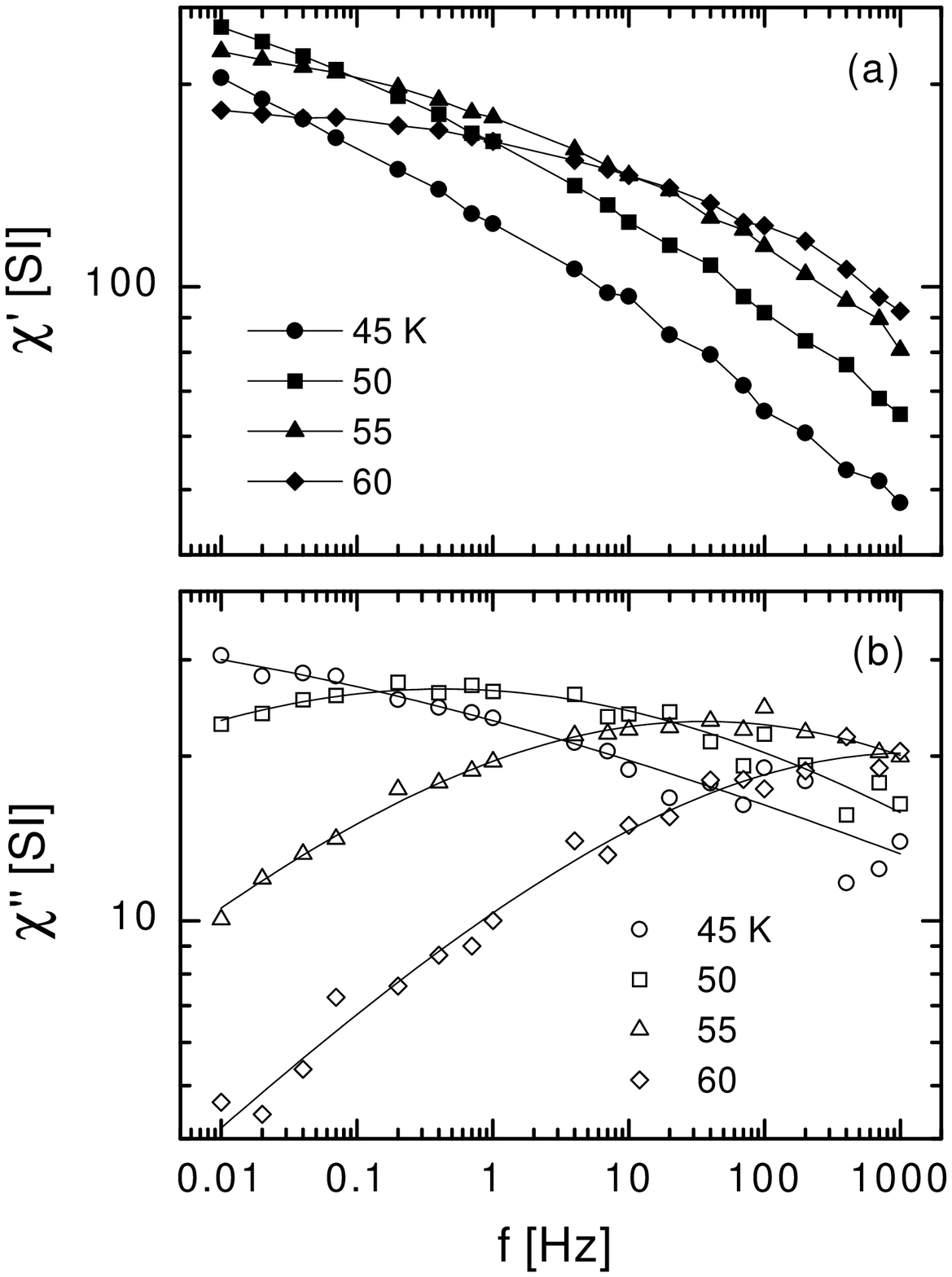, width=13.8cm}
\caption{$\chi'$ (a) and $\chi''$ (b) vs $f$ at 
different temperatures $T =$ 45, 50, 55 and 60~K
for the $t = 0.9$~nm sample in zero-field and
$\mu_0 H_{ac} = 0.05$~mT. The lines in (a) are guides
to the eyes and in (b) best fits according to Eq.~\ref{eq-imag}.
}
\label{fig-spec09}
\end{center} \end{figure}

\begin{figure}[th] \begin{center}
\epsfig{figure=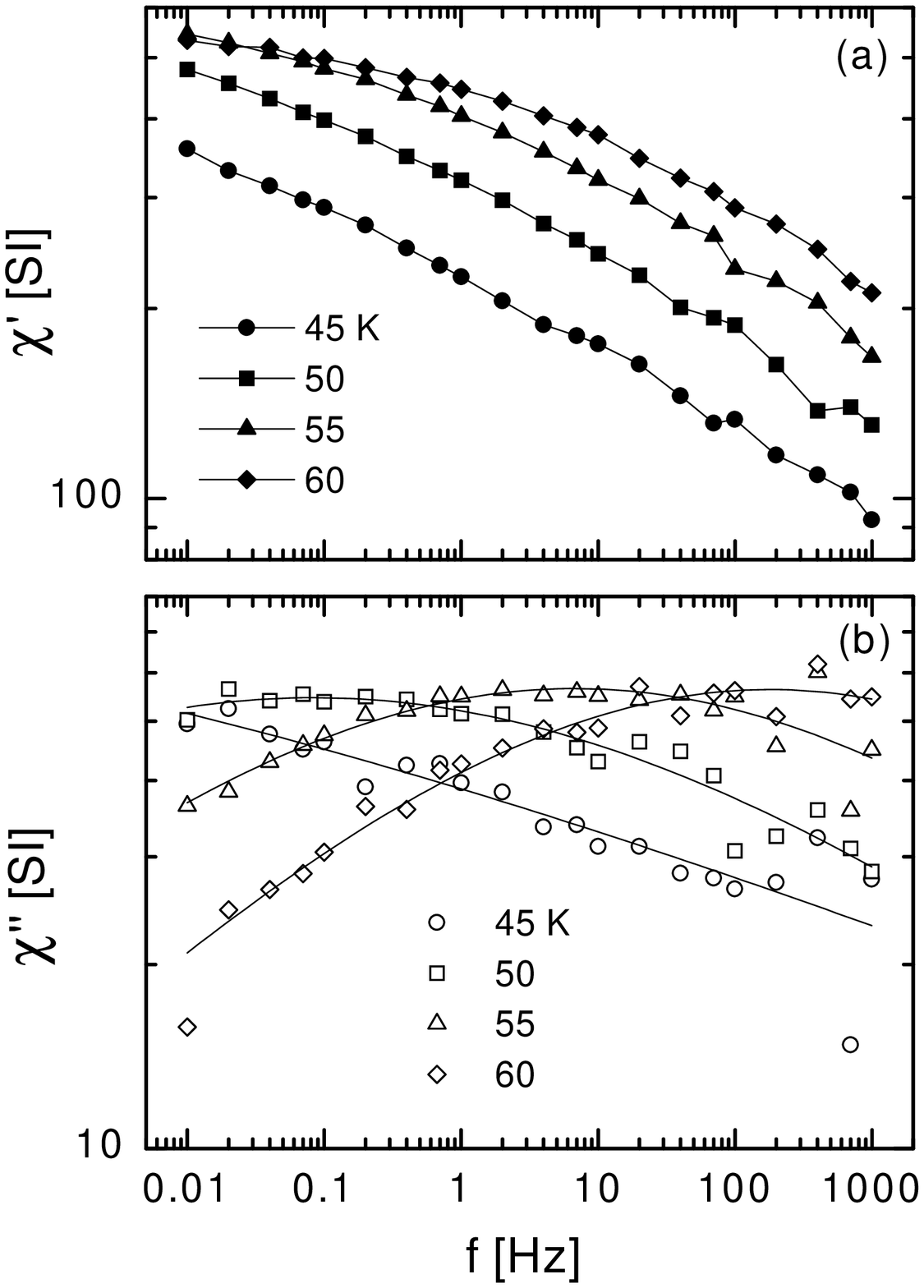, width=13.8cm}
\caption{$\chi'$ (a) and $\chi''$ (b) vs $f$ at 
different temperatures $T =$ 45, 50, 55 and 60~K
for the $t = 1.0$~nm sample in zero-field and
$\mu_0 H_{ac} = 0.05$~mT. The lines in (a) are guides
to the eyes and in (b) best fits according to Eq.~\ref{eq-imag}.
}
\label{fig-spec10}
\end{center} \end{figure}

Next we studied the frequency spectra,
$\chi'(f)$ and $\chi''(f)$. Figure~\ref{fig-spec09}
and Fig.~\ref{fig-spec10} ($t = 0.9$ and $1.0$~nm respectively) show
the real $\chi'$ (a) and the imaginary part $\chi''$ (b) as functions of the
{\it ac} frequency $f$ for different temperatures $T =$ 45, 50, 55 and 60~K
in zero-field and $\mu_0 H_{ac} = 0.05$~mT. 
While some negative curvature still indicates a well-defined dispersion
step at $f > 10^3$~Hz for $T > 60$~K, this step becomes gradually
broadened as $T$ decreases. 
At low $T$ the real parts show
nearly constant negative slopes, 
thus corresponding to an extremely broad
dispersion step. The imaginary parts reveal extremely broad peaks, 
which strongly shift to lower 
frequencies with decreasing temperature. Obviously 
our SSG system exhibits a very wide distribution of relaxation
times with a pronounced temperature dependence. 

A satisfactory
description of the data is provided by the phenomenological Cole-Cole
model \citep*{ColeCole_41, Jonscher_book} 
and was successfully applied e.g. to 2-dimensional (2D) 
\citep*{Dekker_89, Hagiwara_98}
or pseudo-1-dimensional SG systems
\citep*{Ravindran_89}.
The complex {\it ac} susceptibility, $\chi = \chi' - i \chi''$, is written
in the Cole-Cole model as \citep*{Jonscher_book}

\begin{equation} \label{eq-colecole}
\chi(\omega) = \chi_s+ \frac{\chi_0-\chi_s}{1+ (i \omega \tau_c)^{1-\alpha}},
\end{equation}

\noindent where $\chi_0$ and $\chi_S$ are the isothermal 
(low-f) and adiabatic 
(high-f) susceptibilities, respectively, $\tau_c$ is the characteristic
relaxation time and $\alpha$ a measure of the
polydispersivity of the system. The case $\alpha = 0$ yields the 
standard Debye-type relaxator with one single relaxation frequency,
as found, e.g., in the case of a monodisperse ensemble of
non-interacting superparamagnetic particles 
obeying 
Eq.~\ref{eq-neel}. The limiting case $\alpha = 1$ 
corresponds to an infinitely wide distribution of relaxation
times. In SG systems one expects values of $\alpha $ near to 1.

After decomposing Eq.~\ref{eq-colecole} into its real and
imaginary parts 
it is possible to perform a fit to the data 
as shown in Fig.~\ref{fig-spec09} and \ref{fig-spec10}.
One finds 
(compare to Dekker {\it et al.}, 1989; Ravindran {\it et al.}, 1989)

\begin{eqnarray} \label{eq-real}
\chi'(\omega) & = & \chi_S+\frac{\chi_0-\chi_S}{2}
\left( 1-\frac{\sinh \lbrack (1-\alpha) \ln(\omega \tau_c) \rbrack}
{\cosh \lbrack (1-\alpha) \ln(\omega \tau_c) \rbrack + 
\cos \lbrack \frac{1}{2}(1-\alpha)\pi \rbrack } \right) \\ \label{eq-imag}
\chi''(\omega) & = & \frac{\chi_0-\chi_S}{2}
\left( \frac{\sin \lbrack \frac{1}{2}(1-\alpha ) \pi \rbrack }
{\cosh \lbrack (1-\alpha) \ln (\omega \tau_c) \rbrack + 
\cos \lbrack \frac{1}{2}(1-\alpha) \pi \rbrack } \right),
\end{eqnarray}

\noindent where $\omega = 2 \pi f$.
Best results are obtained, when 
fitting to the imaginary part $\chi''(f)$, since only
three parameters, $\chi_0-\chi_S$, $(1-\alpha)$ and $\tau_c$ are
needed in this case.

\begin{figure}[th] \begin{center}
\epsfig{figure=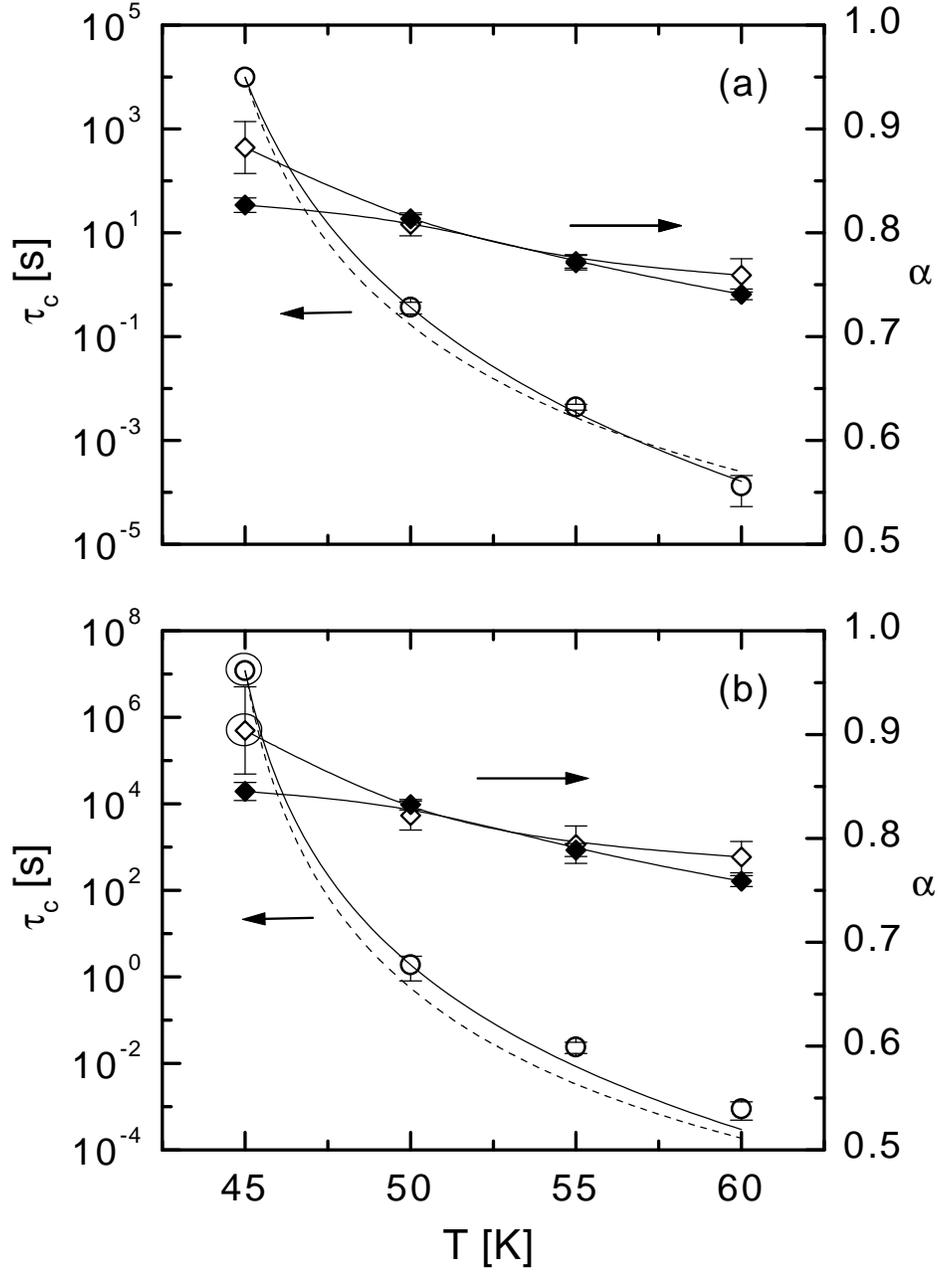, width=13.8cm}
\caption{Results from Cole-Cole fits to the data
shown in Fig.~\ref{fig-spec09}, \ref{fig-spec10} and
\ref{fig-colecole} for
$t=0.9$ (a) and $1.0$~nm (b). 
The characteristic relaxation times $\tau_c$
(open circles) are best fitted to a critical power law 
(solid and broken lines; see text). The 
polydispersivity exponent $\alpha$ vs
$T$ as obtained
from fits to Eq.~\ref{eq-imag} (open
diamonds) and to Eq.~\ref{eq-chichi} (solid diamonds), 
respectively, are connected by eye-guiding lines.  
Encircled data points are results from
non-converging fits.
}
\label{fig-param}
\end{center} \end{figure}

Figure \ref{fig-param} shows the results from the fitting,  
$\tau_c$ (open circles) and $\alpha$ vs $T$ (open diamonds)
for both samples, $t=0.9$ (a) and 
$1.0$~nm (b). One finds that the characteristic
relaxation time $\tau_c$ is increasing with decreasing temperature.
It changes by  eight (a) or ten orders (b) of magnitude, respectively. 
By this kind of extraction of $\tau_c$ one has
access to an extremely wide timescale and is, hence, more 
advantageous compared to the
standard method of extracting $\tau_c$ from the $\chi'(T)$ data.
It is straightforward to perform a fit of the $\tau_c(T)$ data to a 
critical power-law, which was already used in previous
publications 
\citep*{Kleemann_01a, Petracic_02a}, 
$\tau_c = \tau_0 (T/T_g-1)^{-z \nu}$
(solid line). It yields reasonable results, but the value
of $z \nu$ must be kept restricted or even fixed to 
$z \nu = 9$. Then we obtain 
$\tau_0= (5.0\cdot10^{-8} \pm 5.1\cdot10^{-8})$~s, 
$T_g = (42.63 \pm 0.18)$~K and $z \nu=9.0 \pm 0.7$ ($t=0.9$~nm)
and  
$\tau_0= (3.46\cdot10^{-8} \pm 2.1\cdot10^{-10})$~s, 
$T_g = (43.9354 \pm 0.0002)$~K and $z \nu=9$ fixed ($t=1.0$~nm), 
respectively. 
In the case of $t = 0.9$~nm the 
values for $T_g$ and $\tau_0$ correspond well to the
values obtained previously 
\citep*{Kleemann_01a, Petracic_02a}. 
This does not apply to the $t=1.0$~nm sample, where
$T_g \approx 44$~K differs strongly from the value shown
above, $T_g \approx 49$~K. 
Interestingly the
Cole-Cole fit to the $T=45$~K data for $t=1.0$~nm does not
converge (encircled data points in Fig.~\ref{fig-param} (b))
leading to the conclusion that the data emerge from the 
non-ergodic regime, $T < T_g$. 
It is worth to mention that the fit to the modified power law according
to \citep*{Souletie_85}, $\tau_c = \tau_0 (1-T_g/T)^{-z \nu}$, 
(broken line) yields similar values, i.e.  
$\tau_0= (2.67\cdot10^{-9} \pm 2.9\cdot10^{-9})$~s, 
$T_g = (43.19 \pm 0.11)$~K and $z \nu=9.0$ fixed ($t=0.9$~nm)
and  
$\tau_0= (3.4\cdot10^{-10} \pm 6.2\cdot10^{-10})$~s, 
$T_g = (44.004 \pm 0.036)$~K and $z \nu=10$ fixed ($t=1.0$~nm), 
respectively. It is not possible to judge about the 
advantage of this method here. 

The exponent $\alpha$ increases, as expected, with decreasing 
temperature (Fig.~\ref{fig-param}.
Its high value ($\alpha \approx 0.8$) meets the expectation,
that a SG system should have 
a very broad distribution of relaxation times
\citep*{Mydosh_book}.

\begin{figure}[th] \begin{center}
\epsfig{figure=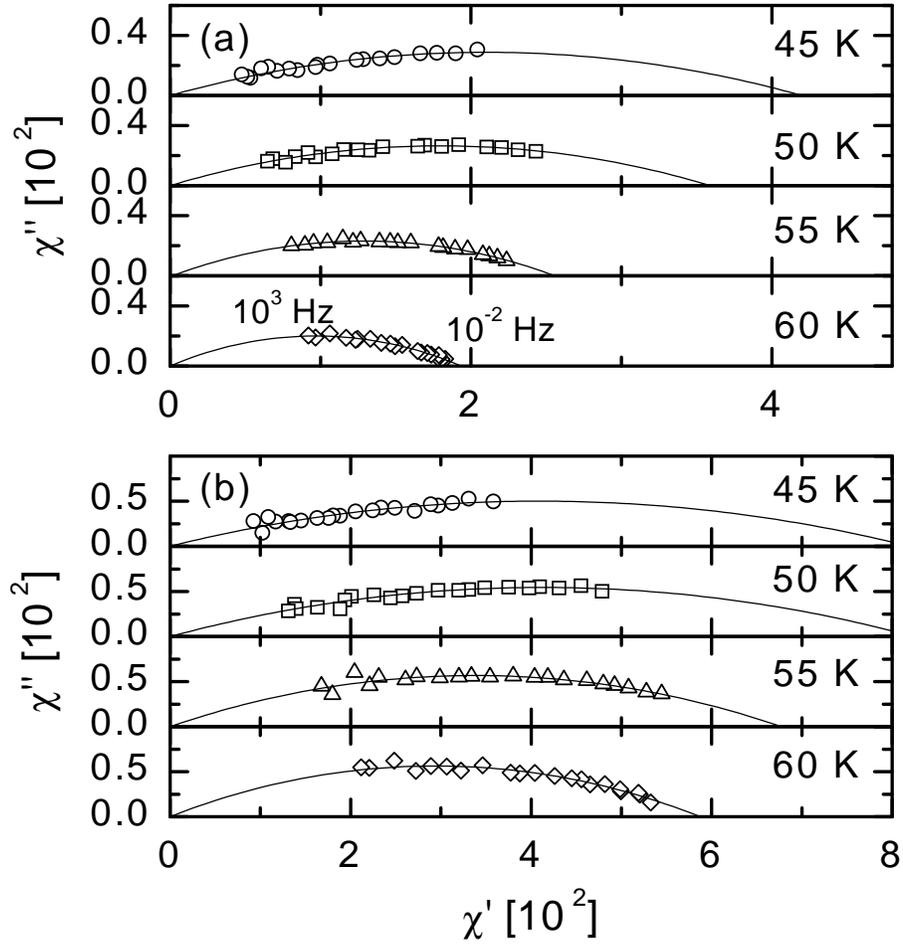, width=13.8cm}
\caption{Cole-Cole plots of $\chi''$ vs $\chi'$ 
for $t=0.9$ (a) and 1.0~nm (b) 
at different temperatures and frequencies as indicated.
The solid lines are best fits according to Eq.~\ref{eq-chichi}.
}
\label{fig-colecole}
\end{center} \end{figure}

Often suscpetibility data are presented in a way, where the imaginary 
part is plotted against the real part (Cole-Cole plot), $\chi''(\chi')$
\citep*{ColeCole_41, Jonscher_book}, where
a classic Debye-relaxator should yield a perfect semicircle, centered
on the $\chi'$-axis at $(\chi_0+\chi_S)/2$ and with radius 
$(\chi_0-\chi_S)/2$. 
The apex of the semi-circle corresponds to $\omega \tau_c = 1$.
Non-zero $\alpha$
has the effect to depress the semi-circle such that the angles between
the $\chi'$-axis and the tangents at $\omega = 0$ and 
$\omega \rightarrow \infty$ are $\mp (1-\alpha) \pi / 2$, respectively.
Figure~\ref{fig-colecole} shows the susceptibility data for $t=0.9$ (a) and
$1.0$~nm at different temperatures $T= 45$, 50, 55 and 60~K.  
The above derived expressions for the real
and imaginary parts (Eq.~\ref{eq-real} and \ref{eq-imag}) can be 
expressed in the form 
\citep*{Hagiwara_98}

\begin{equation} \label{eq-chichi}
\chi''(\chi') = -\frac{\chi_0-\chi_S}
{2 \tan \lbrack (1-\alpha) \pi /2 \rbrack }
+ \sqrt{(\chi'-\chi_S)(\chi_0-\chi')
+\frac{(\chi_0-\chi_S)^2}{4 \tan^2 \lbrack (1-\alpha) \pi /2 \rbrack}}.
\end{equation}

\noindent The fit yields similar results for $\alpha(T)$ 
compared to those from the fit to the imaginary
part $\chi''(f)$ (Fig.~\ref{fig-param}, solid 
versus open diamonds). 
It should be noticed that $\chi_S = 0$ in all cases, i.e. no
measurable response is expected at frequencies above single
particle flip frequencies. This corroborates the model of
ferromagnetic order within each superparamagnetic particle.

\section*{CONCLUSION}

The dynamical susceptibility of the SSG system 
$\lbrack$Co$_{80}$Fe$_{20}(t)$/Al$_{2}$O$_{3}$(3nm)$\rbrack$$_{10}$
($t=0.9$ and 1.0~nm) 
was studied under the 
influence of a bias field and in view of its polydispersivity within the
framework of a Cole-Cole description. 
Cole-Cole fits yield reasonable values for the characteristic
relaxation time $\tau_c$ of the system and for its polydispersivity 
exponent $\alpha$. The relaxation time can be well described by a 
critical power-law dependence. 
One should note that by this kind of extraction of $\tau_c$ one has
access to an extremely wide timescale of eight or ten orders of 
magnitude. Reasonably large values, 
$\alpha \approx 0.8$, are obtained, which are 
typical of SG systems. The Cole-Cole plots of
the susceptibility data  confirm the SG characteristic, i.e. one observes
a strongly flattened semi-circle.
\\*[0.2cm]

\noindent{\bf \em Acknowledgements}
\\

\noindent
The authors acknowledge financial support by DFG
(Graduiertenkolleg "Struktur und Dynamik heterogener Systeme").

{\small
  \bibliography{petracic}

\begin{thebibliography}{25}
\expandafter\ifx\csname natexlab\endcsname\relax\def\natexlab#1{#1}\fi
\expandafter\ifx\csname url\endcsname\relax
  \def\url#1{{\tt #1}}\fi
\expandafter\ifx\csname urlprefix\endcsname\relax\def\urlprefix{URL }\fi

\bibitem[{Barbara {\em et~al.\/}(1981)Barbara, Malozemoff and
  Imry}]{Barbara_81}
Barbara, B., A.~Malozemoff and Y.~Imry (1981).
\newblock Field-dependence of the dc susceptibility of spin glasses.
\newblock {\em Physica B \& C\/} {\bf \bf 108}, 1289.

\bibitem[{Brown(1963)}]{Brown_63}
Brown, W. (1963).
\newblock Thermal fluctuations of a single-domain particle.
\newblock {\em Phys. Rev.\/} {\bf \bf 130}, 1677.

\bibitem[{Canella and Mydosh(1972)}]{Canella_72}
Canella, V. and J.~Mydosh (1972).
\newblock Magnetic ordering in gold-iron alloys.
\newblock {\em Phys. Rev. B\/} {\bf \bf 6}, 4220.

\bibitem[{Chen {\em et~al.\/}(2002)Chen, Sichelschmidt, Kleemann, Petracic {\em
  et~al.\/}}]{Chen_02a}
Chen, X., O.~Sichelschmidt, W.~Kleemann, O.~Petracic {\em et~al.\/} (2002).
\newblock Domain wall relaxation, creep, sliding and switching in
  superferromagnetic discontinuous {C}o$_{80}${F}e$_{20}$-{A}l$_2${O}$_3$
  multilayers.
\newblock {\em Phys. Rev. Lett., submitted\/} .

\bibitem[{Cole and Cole(1941)}]{ColeCole_41}
Cole, K. and R.~Cole (1941).
\newblock Dispersion and absorption in dielectrics.
\newblock {\em J. Chem. Phys.\/} {\bf \bf 9}, 341.

\bibitem[{Dekker {\em et~al.\/}(1989)Dekker, Arts, \mbox{de~Wijn} and
  \mbox{van~Duyneveldt}}]{Dekker_89}
Dekker, C., A.~Arts, H.~\mbox{de~Wijn} and A.~\mbox{van~Duyneveldt} (1989).
\newblock Activated dynamics in a two-dimensional {I}sing spin glass:
  {R}b$_2${C}u$_{1-x}${C}o$_x${F}$_4$.
\newblock {\em Phys. Rev. B\/} {\bf \bf 40}, 11243.

\bibitem[{Djurberg {\em et~al.\/}(1997)Djurberg, Svedlindh, Nordblad, Hansen
  {\em et~al.\/}}]{Djurberg_97}
Djurberg, C., P.~Svedlindh, P.~Nordblad, M.~Hansen {\em et~al.\/} (1997).
\newblock Dynamics of an interacting particle system: evidence of critical
  slowing down.
\newblock {\em Phys. Rev. Lett.\/} {\bf \bf 79}, 5154.

\bibitem[{Dormann {\em et~al.\/}(1999)Dormann, Fiorani, Cherkaoui, Tronc {\em
  et~al.\/}}]{Dormann_99b}
Dormann, J., D.~Fiorani, R.~Cherkaoui, E.~Tronc {\em et~al.\/} (1999).
\newblock From pure superparamagnetism to glass collective state in
  $\gamma$-{F}e$_2${O}$_3$ nanoparticle assemblies.
\newblock {\em J. Magn. Magn. Mater.\/} {\bf \bf 203}, 23.

\bibitem[{Dormann {\em et~al.\/}(1997)Dormann, Fiorani and Tronc}]{Dormann_97}
Dormann, J., D.~Fiorani and E.~Tronc (1997).
\newblock Magnetic relaxation in fine-particle systems.
\newblock {\em Adv. Chem. Phys.\/} {\bf \bf 98}, 283.

\bibitem[{Hagiwara(1998)}]{Hagiwara_98}
Hagiwara, M. (1998).
\newblock Cole-{C}ole plot analysis of the spin-glass system {N}i{C}$_2${O}$_4
  \cdot 2(2${M}{I}z$)_{0.49}(${H}$_2${0}$)_{0.51}$.
\newblock {\em J. Magn. Magn. Mater.\/} {\bf \bf 177-181}, 89.

\bibitem[{Jonscher(1983)}]{Jonscher_book}
Jonscher, A. (1983).
\newblock In {\em Dielectric relaxation in solids\/}. Chelsea Dielectrics
  Press, London.

\bibitem[{Kakazei {\em et~al.\/}(2001)Kakazei, Pogorelov, Lopes, Sousa {\em
  et~al.\/}}]{Kakazei_01}
Kakazei, G., Y.~Pogorelov, A.~Lopes, J.~Sousa {\em et~al.\/} (2001).
\newblock Tunnel magnetoresistance and magnetic ordering in ion-beam sputtered
  {C}o$_{80}${F}e$_{20}$/{A}l$_2${O}$_3$ discontinuous multilayers.
\newblock {\em J. Appl. Phys.\/} {\bf \bf 90}, 4044.

\bibitem[{Kleemann {\em et~al.\/}(2001)Kleemann, Petracic, Binek, Kakazei {\em
  et~al.\/}}]{Kleemann_01a}
Kleemann, W., O.~Petracic, C.~Binek, G.~Kakazei {\em et~al.\/} (2001).
\newblock Interacting ferromagnetic nanoparticles in discontinuous
  {C}o$_{80}${F}e$_{20}$/{A}l$_2${O}$_3$ multilayers: from superspin glass to
  reentrant superferromagnetism.
\newblock {\em Phys. Rev. B\/} {\bf \bf 63}, 134423.

\bibitem[{Mamiya {\em et~al.\/}(1999)Mamiya, Nakatani and
  Furubayashi}]{Mamiya_99}
Mamiya, H., I.~Nakatani and T.~Furubayashi (1999).
\newblock Slow dynamics for spin-glass-like phase of a ferromagnetic fine
  particle system.
\newblock {\em Phys. Rev. Lett.\/} {\bf \bf 82}, 4332.

\bibitem[{Mydosh(1993)}]{Mydosh_book}
Mydosh, J. (1993).
\newblock In {\em Spin glasses: an experimental introduction\/}. Taylor \&
  Francis, London.

\bibitem[{N\'eel(1949)}]{Neel_49}
N\'eel, L. (1949).
\newblock Th\'eorie du trainage magn\'etique des ferromagn\'etiques en grains
  fins avec applications aux terres cuites.
\newblock {\em Ann. Geophys.\/} {\bf \bf 5}, 99.

\bibitem[{Petracic {\em et~al.\/}(2002)Petracic, Kleemann, Binek, Kakazei {\em
  et~al.\/}}]{Petracic_02a}
Petracic, O., W.~Kleemann, C.~Binek, G.~Kakazei {\em et~al.\/} (2002).
\newblock Superspin glass behavior of interacting ferromagnetic nanoparticles
  in discontinuous magnetic multilayers.
\newblock {\em Phase Transitions\/} {\bf \bf 75}, 73.

\bibitem[{Ravindran {\em et~al.\/}(1989)Ravindran, Rubenacker, Haines and
  Drumheller}]{Ravindran_89}
Ravindran, K., G.~Rubenacker, D.~Haines and J.~Drumheller (1989).
\newblock Spin-cluster relaxation times in the spin-glass $\lbrack
  (${C}{H}$_3)_3${N}{H}$\rbrack$ {C}o$_{0.4}$ {N}i$_{0.6}${C}l$_3 \cdot
  $2{H}$_2${O}.
\newblock {\em Phys. Rev. B\/} {\bf \bf 40}, 9431.

\bibitem[{Sahoo {\em et~al.\/}(2002{\natexlab{a}})Sahoo, Petracic, Binek,
  Kleemann {\em et~al.\/}}]{Sahoo_02a}
Sahoo, S., O.~Petracic, C.~Binek, W.~Kleemann {\em et~al.\/}
  (2002{\natexlab{a}}).
\newblock Superspin-glass nature of discontinuous
  {C}o$_{80}${F}e$_{20}$/{A}l$_2${O}$_3$ multilayers.
\newblock {\em Phys. Rev. B\/} {\bf \bf 65}, 134406.

\bibitem[{Sahoo {\em et~al.\/}(2002{\natexlab{b}})Sahoo, Sichelschmidt,
  Petracic, Binek {\em et~al.\/}}]{Sahoo_02b}
Sahoo, S., O.~Sichelschmidt, O.~Petracic, C.~Binek {\em et~al.\/}
  (2002{\natexlab{b}}).
\newblock Magnetic states of discontinuous
  {C}o$_{80}${F}e$_{20}$-{A}l$_2${O}$_3$ multilayers.
\newblock {\em J. Magn. Magn. Mater.\/} {\bf \bf 240}, 433.

\bibitem[{Sankar {\em et~al.\/}(2000)Sankar, Dender, Borchers, Smith {\em
  et~al.\/}}]{Sankar_00}
Sankar, S., D.~Dender, J.~Borchers, D.~Smith {\em et~al.\/} (2000).
\newblock Magnetic correlations in non-percolated {C}o-{S}i{O}$_2$ granular
  films.
\newblock {\em J. Magn. Magn. Mater.\/} {\bf \bf 221}, 1.

\bibitem[{Souletie and Tholence(1985)}]{Souletie_85}
Souletie, J. and J.~Tholence (1985).
\newblock Critical slowing down in spin glasses and other glasses : {F}ulcher
  versus power law.
\newblock {\em Phys. Rev. B\/} {\bf \bf 32}, 516.

\bibitem[{Sousa {\em et~al.\/}(2001)Sousa, Kakazei, Pogorelov, Santos {\em
  et~al.\/}}]{Sousa_01}
Sousa, J., G.~Kakazei, Y.~Pogorelov, J.~Santos {\em et~al.\/} (2001).
\newblock Magnetic states of granular layered {C}o{F}e-{A}l$_2${O}$_3$ system.
\newblock {\em IEEE Trans. Mag.\/} {\bf \bf 37}, 2200.

\bibitem[{Stappert {\em et~al.\/}(2002)Stappert, Dumpich, Sahoo, Petracic {\em
  et~al.\/}}]{Stappert_02}
Stappert, S., G.~Dumpich, S.~Sahoo, O.~Petracic {\em et~al.\/} (2002).
\newblock Transmission electron microscopy studies on ion-beam sputtered
  {C}o$_{80}${F}e$_{20}$/{A}l$_2${O}$_3$ discontinuous bilayers.
\newblock {\em unpublished\/} .

\bibitem[{Young(1997)}]{Young_book}
Young, A. (1997).
\newblock ({E}ditor), {S}pin glasses and random fields.
\newblock In {\em Series on directions in condensed matter physics\/},  Vol.
  12. World Scientific, Singapore.

\end{thebibliography}
  \bibliographystyle{phase-trans}
}

\end{document}